\documentclass[12pt]{article}
\usepackage{epsf}
\def\be{\begin{equation}}
\def\ee{\end{equation}}
\def\bea{\begin{eqnarray}}
\def\eea{\end{eqnarray}}

\begin{document}
\begin{titlepage}
\begin{center}
{\Large \bf William I. Fine Theoretical Physics Institute \\
University of Minnesota \\}
\end{center}
\vspace{0.2in}
\begin{flushright}
FTPI-MINN-16/15 \\
UMN-TH-3525/16 \\
April 2016 \\
\end{flushright}
\vspace{0.3in}
\begin{center}
{\Large \bf Sum rules for interaction of $\Upsilon$ resonances with $Z_b \pi$ 
\\}
\vspace{0.2in}
{\bf  M.B. Voloshin  \\ }
William I. Fine Theoretical Physics Institute, University of
Minnesota,\\ Minneapolis, MN 55455, USA \\
School of Physics and Astronomy, University of Minnesota, Minneapolis, MN 55455, USA \\ and \\
Institute of Theoretical and Experimental Physics, Moscow, 117218, Russia
\\[0.2in]

\end{center}

\vspace{0.2in}

\begin{abstract}
The strength of the amplitudes for the coupling between the bottomonium $\Upsilon(nS)$ states, the bottomonium-like isovector resonances $Z_b$ and a pion, $\Upsilon(nS) Z_b \pi$, is considered. These amplitudes describe the decays $Z_b \to \Upsilon(nS) \pi$ for $n=1,\,2,\,3$, and the processes $\Upsilon(nS) \to Z_b \pi$ for $n=5,\,6, \ldots$ with either $Z_b(10610)$ or $Z_b(10650)$. It is pointed out that analyticity and unitarity impose a sum rule for these couplings to each of the $Z_b$ resonances. With the currently available data it appears to be difficult or impossible to simultaneously satisfy the sum rules for the $Z_b(10610)$ and $Z_b(10650)$ resonances. This difficulty can be resolved if there is a considerable dissimilarity in the yield of the states $Z_b(10610) \pi$ and  $Z_b(10650) \pi$ in the $e^+e^-$ annihilation at energies above the $\Upsilon(5S)$ resonance.

\end{abstract}
\end{titlepage}

The bottomonium-like isovector resonances~\cite{bellez,bellez0} $Z_b(10610)$ and $Z_b(10650)$ with the quantum numbers $I^G(J^P) =1^+(1^+)$ are objects of great interest for experimental and theoretical studies due to their manifestly exotic quark structure. Their masses, $10607.2 \pm 2.0\,$MeV and $10652.2 \pm 1.5\,$MeV coincide, within the errors, with the thresholds for respectively $B^* \bar B$ and $B^* \bar B^*$ heavy meson pairs, suggesting~\cite{bgmmv} that these are `molecular' states~\cite{ov} of the corresponding heavy meson-antimeson pair.
These resonances are mostly studied through their production in the $e^+e^-$ annihilation at the c.m. energy around 10870\,MeV within the peak of the bottomonium resonance $\Upsilon(5S)$~\cite{belle1403,belle1512} due to the decay $\Upsilon(5S) \to Z_b \pi$, and there is also an indication~\cite{belle1501,belle1508} of similar processes at higher energies around 11000\,MeV corresponding to the peak of the $\Upsilon(6S)$ resonance. The $Z_b$ resonances are observed through their decay into lower bottomonium states and a pion: $Z_b \to \Upsilon(nS) \pi$ with $n=1,\,2,\,3$ and $Z_b \to h_b(kP) \pi$ with $k=1,\,2$~\cite{belle1403}, or into the corresponding heavy meson pairs: $Z_b(10610) \to B^* \bar B + c.c.$,  $Z_b(10650) \to B^* \bar B^*$~\cite{belle1512}.

The purpose of this paper is to point out that there is a sum rule, following from analyticity and unitarity, and relating the couplings of bottomonium states with fixed quantum numbers $J^{PC}$ to the channel $Z_b \pi$ with each one of the $Z_b$ resonances. The sum runs over the products of these couplings times the amplitude of production of the bottomonium states by a local operator producing the $b \bar b$ quark pair with the corresponding quantum numbers. Clearly, of the most practical interest are the sum rules for the vector $1^{--}$ bottomonium states, $\Upsilon(nS)$ resonances, since their production by the electromagnetic current of the $b$ quarks, $j_\mu= (\bar b \gamma_\mu b)$, is directly measured in the $e^+e^-$ annihilation, and the discussed sum rules can be compared with the existing data. For this reason the further discussion is presented for this channel. 

It will be argued below that for each of the $Z_b$ resonances there are two sum rules relating the couplings between $Z_b \pi$ and the $\Upsilon(nS)$ resonances. The two sum rules differ by the power of the masses of the $\Upsilon(nS)$. The sums run over all the vector $b \bar b$ resonances including the contribution of the $\Upsilon(4S)$. However the strength of the coupling between $Z_b \pi$ and the latter state cannot be directly measured for kinematical reasons, and in order to eliminate this unknown contribution the two sum rules are combined into one relation having the form 
\be
\sum_n \, C_n (M_4-M_n) + {\rm continuum} = 0~,
\label{sr}
\ee
where the sum includes all the $\Upsilon(nS)$ resonances with masses $M_n$, and the quantities $C_n$ are generally complex and their absolute values squared are related to the $e^+ e^-$ decay widths, $\Gamma_{ee}$ of these resonances and the rates of the one-pion transitions between these and one of the $Z_b$ resonances as
\be
|C_n|^2 =  {\rm const} \, {\Gamma_{ee}[\Upsilon(nS)] \, \Gamma_{Z\pi} [\Upsilon(nS)] \over E_\pi^2 k_\pi}~.
\label{cn2}
\ee
Here $\Gamma_{Z\pi} [\Upsilon(nS)]$ stands for the rate of the kinematically possible transition: $\Gamma[Z_b \to \Upsilon(nS)]$ for $n=1,\,2,\,3$ and $\Gamma[\Upsilon(nS) \to Z_b \pi]$ for $n=5,\,6,\ldots$; $E_\pi$ and $k_\pi$ denote the energy and the momentum of the emitted pion. Finally, the `continuum' term is a contribution of the production of $Z_b \pi$ states in $e^+e^-$ in the continuum, i.e. not associated with the bottomonium resonances. The common constant in Eq.(\ref{cn2}), inessential in  the relation (\ref{sr}), can be chosen for `numerical convenience', and will be specified later in the text.

A certain deficiency of the sum rule (\ref{sr}) is that the terms in it are not sign-definite but rather are generally complex numbers, $C_n= e^{i \phi_n} |C_n|$ , and only the absolute value of the contribution of each $\Upsilon$ resonance can be evaluated from Eq.(\ref{cn2}) using the data. However the phases between the coefficients $C_n$ for the $Z_b(10610)$ and $Z_b(10650)$ (i.e., in different sum rules) for each of the lower $\Upsilon(nS)$ ($n=1,\,2,\,3$) states can be determined from the interference patterns between the two $Z_b$ resonances in the corresponding channel~\cite{belle1403}. 

Furthermore, neither the continuum contribution has a definite phase.
At the energies up to 11020\,MeV studied thus far, no non-resonant production of the $Z_b \pi$ states has been observed~\cite{belle1508}, so that the continuum term is very small or zero. If it is found that with the studied $\Upsilon(nS)$ up to $n=6$ the sum rules can not be satisfied for one or both $Z_b$ resonances, this would imply existence of their production at higher $e^+e^-$ energies, either in the continuum, or in possible higher resonances. 

In fact, even with the current uncertainty, it can be concluded that limiting the summation in Eq.(\ref{sr}) at the $\Upsilon(5S)$ resonance indeed results in a significant mismatch between the sums for the $Z_b(10610)$ and $Z_b(10650)$ states. As will be discussed, the absolute values of the coefficients $C_n$ with $n=1,\,2,\,3$ in the sum rule for $Z_b(10650)$ are systematically smaller (by approximately a factor of two) than those in the sum rule for $Z_b(10610)$. On the contrary,
the absolute values of the coefficients $C_5$ in the sum rules for each of these states are very close to each other, although the overall normalization is subject to an uncertainty. For this reason it is impossible to saturate the sum rules for both $Z_b(10610)$ and $Z_b(10650)$ simultaneously by adjusting the common normalization of the coefficients $C_5$. This considerable mismatch should thus be compensated by a significantly dissimilar yield in the channels $Z_b(10610) \pi$  and $Z_b(10650) \pi$ at $\Upsilon(6S)$ or at higher energies~\footnote{The study in Ref.~\cite{belle1508} could not resolve the relative yield in these channels at $\Upsilon(6S)$, and it has been also argued in a model~\cite{bv}, that this relative yield can display a nontrivial behavior at $\Upsilon(6S)$ and at a higher energy around 11.2\,GeV.}.

The derivation of the sum rules is based on considering the amplitude for production of the state $Z_b \pi$ with one of the $Z_b$ bosons by the $b$ quark electromagnetic current:
\be
A_\mu = \langle Z_b(\epsilon, p) \pi(k) | j_\mu(q) | 0 \rangle~,
\label{aa}
\ee
where $\epsilon$ and $p$ are the polarization amplitude and the momentum of the $Z_b$, and $q=p+k$. The amplitude can be written in terms of two scalar functions, in agreement with the presence of $S$ and $D$ partial waves. The expression,  satisfying  the  relations $(q \cdot A) =0$ and  $(p \cdot \epsilon) = 0$, can generally be written in terms of two invariant form factors $A_1$ and $A_2$,
\be
A_\mu = A_1 \, \left[ \epsilon_\mu \, (q \cdot k) - k_\mu \, (\epsilon \cdot k) \right ] + A_2 \, (\epsilon \cdot k) \, \left[ k_\mu - {(q \cdot k) \over (q \cdot p)} \, p_\mu \right ]~.
\label{a12}
\ee
It can be noted, that the amplitude also automatically satisfies the chiral algebra requirement of vanishing at zero pion four-momentum, so that $A_1$ and $A_2$ are finite at $k \to 0$. The form factors $A_1$ and $A_2$ are functions of three invariants: $q^2,\,p^2$ and $k^2$. For on-shell $Z_b$ and pion, two of the latter invariants are obviously fixed, $p^2=M_Z^2$, $k^2=m_\pi^2$, so that the form factors are analytic functions of $q^2$, $A_{1,2}(q^2)$. The asymptotic behavior of these functions at large $q^2$ is limited by the quark counting rule~\cite{mmt,bf}, according to which they decrease  as $1/|q^2|^{5/2}$ at $|q^2| \to \infty$. One can thus write a dispersion relation 
\be
A_{1,2}(q^2) = {1 \over 2 \, \pi \, i} \, \int \, {{\rm Disc} A_{1,2}(s) \over s- q^2 - i 0} \,ds~,
\label{dr}
\ee
and conclude that the conditions for the required asymptotic behavior at large $|q^2|$ reads
\be
\int \, {\rm Disc} A_{1,2}(s) \, ds =0~,~~~{\rm and}~~\int \, {\rm Disc} A_{1,2}(s) \, s \, ds =0~.
\label{sri}
\ee
The integral runs over the values of $s$ where the discontinuity, Disc$A$, of the amplitude at the unitary cut is nonzero. By the unitarity relation these values correspond to the on-shell vector $b \bar b$ states produced by the electromagnetic current and coupled to $Z_b \pi$. Thus the integral in Eq.(\ref{sri}) is contributed by the  $\Upsilon(nS)$ resonances and, possibly, a continuum at higher energies. In the mass region of the $\Upsilon$ resonances in the discussed pion transitions between bottomonium and the $Z_b$ resonances the hidden-bottom states (but not the pion) can be treated nonrelativistically. Also in these transitions  the $D$ wave is suppressed by the heavy quark spin symmetry (HQSS) and also kinematically, and one can limit consideration to only the $S$ wave, which is given by the first term in the part of the expression (\ref{a12}) proportional to $A_1$. In the nonrelativistic limit the amplitude $A_\mu$ reduces in the center of mass frame to
\be
{\vec A}= A_1 \, M \, {\vec \epsilon} \, E_\pi~,
\label{nra}
\ee 
where $M$ is  hidden-bottom mass which is to be taken as a common constant, since its difference for different states is beyond the leading nonrelativistic order. 

The contribution of each of the $\Upsilon(nS)$ resonances in the imaginary part of $A_1$ can be written as 
\be
\left . {1 \over 2 \, \pi \, i} \, {\rm Disc}A_1 \right |_{\Upsilon(nS)} =  \delta(s-M_n^2) \, M \, C_n~,
\label{acr}
\ee
where for all $n$, except for $n=4$,
\be
|C_n| = \left (  {24 \pi \over Q_b^2 \, \alpha^2} \, {\Gamma_{ee}[\Upsilon(nS)] \, \Gamma_{Z\pi} [\Upsilon(nS)] \, M \over E_\pi^2 k_\pi} \right )^{1/2}
\label{cnm}
\ee
with $Q_b =-1/3$ being the electric charge of the $b$ quark and $\alpha$ the fine structure constant. This relation also specifies, for definiteness, the convention for the overall normalization of the coefficients $C_n$, used in the numerical estimates below. (Also the value $M=10\,$GeV is used for definiteness. Clearly, the relative values of the coefficients $C_n$ do not depend on this specific number.)

The coupling $\Upsilon(4S) Z^{(')}_b \pi$ is not accessible kinematically for a measurement  either in the production of the $Z^{(')}_b \pi$ channel or in the decays of the $Z^{(')}_b$ resonances. In order to eliminate the unknown contribution of the $\Upsilon(4S)$ one can combine the sum rules (\ref{sri}) into one with the weight factor vanishing at the $\Upsilon(4S)$ pole:
\be
\int \, {\rm Disc} A_{1,2}(s) \,(M^2_4 - s) \, ds =0~.
\label{sr4}
\ee
Clearly, in the nonrelativistic limit, where the mass differences between the $\Upsilon(nS)$ resonances are small compared to their common mass $M$, the latter relation reduces to the sum rule in Eq.(\ref{sr}).

The numerical estimates of the absolute values of the coefficients $C_n \equiv C_n[Z_b(10610)]$ and $C_n' \equiv C_n[Z_b(10650)]$ for the lower $\Upsilon(nS)$ resonances with $n=1,\,2,\,3$ can be done using the data~\cite{belle1512} on the branching fractions for the $Z_b$ decays to $\Upsilon(nS) \pi$ and the PDG~\cite{pdg} values of the $\Upsilon(nS)$ leptonic widths $\Gamma_{ee}$ and the widths of the $Z_b$ resonances. The results are shown in Table~1 where in the estimates of the uncertainty the statistical and systematic errors from Ref.~\cite{belle1512} are added in quadrature. The presented numbers are for the amplitudes with one specific charge combination with a charged pion, e.g. $Z_b^+   \to \Upsilon(nS) \pi^+$.

The phases of the coefficients estimated in Table~1 are not known. On the theoretical side, these coefficients can be expected to be almost real, since the complex phase arising from re-scattering in the $\Upsilon(nS) \pi$ channel should be small, as can be deduced from the relatively low rates of similar processes of transition between the lower bottomonium states with emission of two pions. This however leaves an ambiguity in the relative sign of the coefficients $C_n$. 
On the experimental side,
the data on the interference pattern between the $Z_b(10610)$ and $Z_b(10650)$ resonances in the processes $e^+e^- \to \Upsilon(nS) \pi^+ \pi^-$ for each $n$ indicate that the relative phase between their contribution is consistent with zero~\cite{belle1403} (cf. Table~1). 
The experimental observation implies that, given that the absolute value of each of the coefficients $C_n'$ is significantly smaller (by about a factor of two) than of the corresponding coefficient $C_n$, the absolute value of the sum $C_1'+C_2'+C_3'$ should also be similarly smaller than the absolute value of $C_1+C_2+C_3$.
\begin{table}[t]
\caption{The absolute values of the coefficients $C_n \equiv C_n[Z_b(10610)]$ and $C_n' \equiv C_n[Z_b(10650)]$, the products $C_n \, (M_4 - M_n)$ and $C_n' \, (M_4 - M_n)$, and the experimental~\cite{belle1403} relative phases $\phi_n' - \phi_n$ for the lower $\Upsilon(nS)$ resonances.}
\label{tab1}
\vspace{1.5mm}
\begin{tabular}{|c|c|c|c|c|c|}
\hline
n &  $|C_n| $ & $|C_n'|$ & $|C_n|  (M_4 - M_n)$ (MeV) & $|C_n'|  (M_4 - M_n)$ (MeV) & $\phi_n' - \phi_n$ (deg.) \\
\hline
1 & $0.11 \pm 0.02$  & $0.04 \pm 0.02$ & $123 \pm 22$ & $45 \pm 22$ & $67 \pm 36^{+24}_{-52}$ \\
2 & $0.55 \pm 0.08$ & $0.23 \pm 0.04$ & $306 \pm 45$ & $128 \pm 22$ & $-10 \pm 13^{+34}_{-12}$ \\
3 & $1.37 \pm 0.23$ & $0.67 \pm 0.14$ & $308 \pm 52$ &  $152 \pm 32$ & $-5 \pm 22^{+15}_{-33}$ \\
\hline
\end{tabular}
\end{table} 

In order to estimate the coefficients $C_5$ and $C_5'$ describing the amplitudes of the transitions $\Upsilon(5S) \to Z_b(10610) \pi$ and $\Upsilon(5S) \to Z_b(10650) \pi$, one can use the data~\cite{belle1512} on the cross sections $\sigma(E_0) = \sigma[e^+e^- \to (B^* \bar B + \bar B^* B)^+ \pi^-]$ and $\sigma'(E_0)=\sigma[e^+e^- \to (B^* \bar B^*)^+ \pi^-]$ at the energy $E_0=10866\,$MeV where the largest experimental statistics within the peak of $\Upsilon(5S)$ is available~\footnote{I thank A.~Bondar for suggesting this way of estimating the relevant combination of the parameters. Another method leading to similar results, albeit with larger uncertainties, would be based on the measured cross section for $e^+e^- \to h_b(kP) \pi^+ \pi^-$ and the fit~\cite{belle1403} of the fractional contribution of each of the $Z_b$ resonances.}. These data can be converted to the cross section for the production of the $Z_b \pi$ states, since experimentally~\cite{belle1512} the yield of the final states $BB^* \pi$ is only due to the $Z_b(10610)$ resonance and that of $B^*B^*\pi$ is fully described by the $Z_b(10650)$, and the branching fractions are known and measured as $ B_{ZBB^*} = Br[Z_b^+(10610) \to (BB^*)^+]= (82.6 \pm 2.9 \pm 2.3)\%$ and  $ B_{Z'B^*B^*} = Br[Z_b^+(10650) \to (B^*B^*)^+]= (70.6 \pm 4.9 \pm 4.4)\%$. Using the Breit-Wigner formula for the $\Upsilon(5S)$ resonance, one can write the product of its widths $\Gamma_{ee} \Gamma_{Z \pi}$ directly in terms of the measured cross section and the total width $\Gamma_{tot}(5S)$:
\be
\Gamma_{ee}(5S) \Gamma_{Z^{(')} \pi}(5S) = {M_5^2 \, \sigma^{(')}(E_0) \over 12 \pi} \, {\sigma_{max} \over \sigma^{(')}(E_0)} 
{\Gamma_{tot}^2(5S) \over B_{Z^{(')} BB}}~,
\label{bw}
\ee
where $M_5$ is the mass of $\Upsilon(5S)$, and  $\sigma_{max}$ is the cross section at the maximum of the $\Upsilon(5S)$ resonance peak, so that the ratio $r_\sigma(E_0) = \sigma_{max}/\sigma(E_0)$ depends only on the shift of the energy $E_0$ from the resonance maximum at $M_5$:
\be
r_\sigma(E_0) = {4 \, (E_0-M_5)^2 + \Gamma_{tot}^2 \over \Gamma_{tot}^2}~,
\label{bwr}
\ee
and is the same for the $Z_b(10610)$ and $Z_b(10650)$ resonances. The ratio $r_\sigma$ is currently not known well. The data~\cite{belle1508} of the energy scan of the cross section for $e^+e^- \to h_b(kP) \pi^+ \pi^-$, the processes going through the $Z_b$ resonances, indicate that  $r_\sigma(E_0) \approx 2$, and the resonance maximum is actually above 10866\,MeV. In the following numerical estimates a `benchmark' value $r_\sigma(E_0) = 2$ is used. 

The absolute values of the coefficients $C_5$ and $C_5'$ normalized as in Eq.(\ref{cnm}) can thus be evaluated as
\bea
&&|C_5|= (2.10 \pm 0.05)  \,  {\Gamma_{tot}(5S) \over 55\,{\rm MeV}}  \, \sqrt{{r_\sigma(E_0) \over 2} \,  {\sigma(E_0) \over 8.7\,{\rm pb}} }~, \nonumber \\
&&|C_5'| = (2.24 \pm 0.10)  \,  {\Gamma_{tot}(5S) \over 55\,{\rm MeV}}  \, \sqrt{{r_\sigma(E_0) \over 2} \,  {\sigma'(E_0) \over 4.38 \,{\rm pb}} }~,
\label{c5n}
\eea
where the used benchmark values for the cross sections are the central values of the available data~\cite{belle1512} after radiative corrections. (The data presented in Ref.~\cite{belle1512} describe the total cross section for two charge combinations containing a charged pion. Here the cross sections are for one charge combination and are thus two times smaller.) The estimates (\ref{c5n}) translate into the following evaluation of the $\Upsilon(5S)$ contribution to the sums rules (\ref{sr}) for the $Z_b$ and $Z_b'$ resonances
\bea
&&|C_5| (M_4-M_5) = -(630 \pm 28)\,{\rm MeV}  \,  {\Gamma_{tot}(5S) \over 55\,{\rm MeV}}  \, \sqrt{{r_\sigma(E_0) \over 2} \,  {\sigma(E_0) \over 8.7\,{\rm pb}} }~, \nonumber \\
&&|C_5'| (M_4-M_5)= - (672 \pm 39)\,{\rm MeV}  \,  {\Gamma_{tot}(5S) \over 55\,{\rm MeV}}  \, \sqrt{{r_\sigma(E_0) \over 2} \,  {\sigma'(E_0) \over 4.38 \,{\rm pb}} }~,
\label{c5m}
\eea
where the error in the mass difference $M_5-M_4$ is also added in quadrature.

There is a considerable uncertainty in the estimates in Eqs. (\ref{c5n}) and (\ref{c5m}) with a large part of it resulting from the current error~\cite{pdg} in the total width: $\Gamma_{tot}(5S) = 55 \pm 28\,$MeV. The uncertainty is significantly reduced if one considers the ratio of the absolute values of the coefficients:
\be
{|C_5'| \over |C_5|} = (1.07 \pm 0.05) \, \sqrt{ 2 \, \sigma'(E_0) / \sigma(E_0)} = 1.07 \pm 0.13~.
\label{c5r}
\ee
The essentially equal estimated absolute values of these coefficients, combined with the previous estimate of the relative value of the sums over three lower resonances, imply that the sum rules (\ref{sr}) can not be satisfied simultaneously for the $Z_b(10610)$ and $Z_b(10650)$ resonances by only the vector states of bottomonium up to (and including) $\Upsilon(5S)$, and there should be a significant contribution from higher vector $b \bar b$ states. Moreover, the yield of $Z_b(10650) \pi$ at higher energies should be substantially different than that of $Z_b(10610) \pi$. The production of $Z_b \pi$ states has been observed~\cite{belle1501,belle1508} within the $\Upsilon(6S)$ peak. However the ratio of the yield for the two $Z_b$ resonances is not known yet. Also the present uncertainty in the relative amplitude and the phase between the $\Upsilon(5S)$ and $\Upsilon(6S)$ is large, so that it would be premature to speculate whether the production amplitudes at the $\Upsilon(6S)$ can fix the discussed mismatch in the sum rules, or a contribution of still higher states is needed. If however no significant difference in the yield of $Z_b(10650) \pi$ and $Z_b(10610) \pi$ channels is observed in future studies at $\Upsilon(6S)$ and higher energies, the remaining possibility for balancing the sum rules (\ref{sr}) would be that the data on the $Z_b$ resonances should change. Indeed, the suppression of the sum of the coefficients $C_n'$ results, in part, from the smaller measured total width of $Z_b(10650)$ in comparison with $Z_b(10610)$. The tension in the sum rules would thus be somewhat relaxed if this measurement changes in future more detailed data.

In summary. The amplitudes for pion transitions between the $Z_b$ resonances and the bottomonium states $\Upsilon(nS)$ should satisfy the sum rules (\ref{sr}). The existing data indicate that the sums up to $\Upsilon(5S)$ are significantly different for the $Z_b(10610)$ and $Z_b(10650)$ states, and, provided the current data do not change much in the future, the difference should be compensated by a substantially dissimilar yield of the exotic bottomonium-like resonances at $\Upsilon(6S)$ and possibly at higher energies in $e^+e^-$ annihilation. 

I thank Alexander Bondar for illuminating discussions and Bastian Kubis for pointing out an omission in an earlier version of this paper.  This work is supported in part by U.S. Department of Energy Grant No.\ DE-SC0011842.

\end{document}